\documentclass[a4paper,11pt]{article}
\usepackage{graphicx, rotating}
\usepackage{ifpdf}
\ifpdf
\usepackage{hyperref, pdfsync, epstopdf}	
\else
\usepackage[dvips,bookmarks]{hyperref}	
\fi
\hypersetup{colorlinks,bookmarksopen,bookmarksnumbered,citecolor=verdes,
linkcolor=blus,pdfstartview=FitH,urlcolor=rossos}
\def\hhref#1{\href{http://arxiv.org/abs/#1}{#1}} 

\usepackage{multicol}
\usepackage{color}
\definecolor{rosso}{cmyk}{0,1,1,0.4}
\definecolor{rossos}{cmyk}{0,1,1,0.55}
\definecolor{rossoc}{cmyk}{0,1,1,0.2}
\definecolor{blu}{cmyk}{1,1,0,0.3}
\definecolor{blus}{cmyk}{1,1,0,0.6}
\definecolor{bluc}{cmyk}{1,1,0,0.1}
\definecolor{verde}{cmyk}{0.92,0,0.59,0.25}
\definecolor{verdec}{cmyk}{0.92,0,0.59,0.15}
\definecolor{verdes}{cmyk}{0.92,0,0.59,0.4}

\font\tenrsfs=rsfs10 at 11pt
\font\sevenrsfs=rsfs7
\font\fiversfs=rsfs5
\newfam\rsfsfam
\textfont\rsfsfam=\tenrsfs
\scriptfont\rsfsfam=\sevenrsfs
\scriptscriptfont\rsfsfam=\fiversfs
\def\mathscr#1{{\fam\rsfsfam\relax#1}}
\def\Lag{\mathscr{L}}

\newcommand{\fig}[1]{~\ref{fig:#1}}
\newcommand{\eq}[1]{~{\rm (\ref{eq:#1})}}

\newcommand{\eV}{\,{\rm eV}}

\def\circa#1{\,\raise.3ex\hbox{$#1$\kern-.75em\lower1ex\hbox{$\sim$}}\,}

\newcommand{\ds}{\partial\hspace{-1.2ex}/\hspace{0.3ex}}

\newcommand{\PR}{Phys. Rev.}

\newcommand{\beq}{\begin{equation}}
\newcommand{\eeq}{\end{equation}}

\def\circa#1{\,\raise.3ex\hbox{$#1$\kern-.75em\lower1ex\hbox{$\sim$}}\,}
\makeatletter

\def\art{\@ifnextchar[{\eart}{\oart}}
\def\eart[#1]#2#3#4#5#6{{\rm #2}, {#3 #4} {\rm (#6) #5} [{\hhref{#1}}]}
\def\hepart[#1]#2{{\rm #2, \hhref{#1}}}
\newcommand{\oart}[5]{{\rm #1}, {#2 #3} {\rm (#5) #4}}

\newcounter{alphaequation}[equation]
\def\thealphaequation{\theequation\hbox to
0.6em{\hfil\alph{alphaequation}\hfil}}
\def\eqnsystem#1{
\def\@eqnnum{{\rm (\thealphaequation)}}
\def\@@eqncr{\let\@tempa\relax \ifcase\@eqcnt \def\@tempa{& & &} \or
  \def\@tempa{& &}\or \def\@tempa{&}\fi\@tempa
  \if@eqnsw\@eqnnum\refstepcounter{alphaequation}\fi
\global\@eqnswtrue\global\@eqcnt=0\cr}
\refstepcounter{equation} \let\@currentlabel\theequation \def\@tempb{#1}
\ifx\@tempb\empty\else\label{#1}\fi
\refstepcounter{alphaequation}
\let\@currentlabel\thealphaequation
\global\@eqnswtrue\global\@eqcnt=0 \tabskip\@centering\let\\=\@eqncr
$$\halign to \displaywidth\bgroup \@eqnsel\hskip\@centering
$\displaystyle\tabskip\z@{##}$&\global\@eqcnt\@ne
\hskip2\arraycolsep\hfil${##}$\hfil& \global\@eqcnt\tw@\hskip2\arraycolsep
$\displaystyle\tabskip\z@{##}$\hfil
\tabskip\@centering&\llap{##}\tabskip\z@\cr}
\def\endeqnsystem{\@@eqncr\egroup$$\global\@ignoretrue} \makeatother

\newcommand{\SU}{\rm SU}

\begin{document}
\begin{center}
\color{black}
{\Huge\bf\color{rossos}
Cosmology of neutrinos and\\[2mm] extra light particles after WMAP3}
\medskip
\bigskip\color{black}\vspace{0.5cm}

{
{\large\bf Marco Cirelli}$^a$,
{\large\bf Alessandro Strumia}$^b$.
}
\\[7mm]
{\it $^a$ Physics Department, Yale University, New Haven, CT 06520, USA}\\[3mm]
{\it $^b$ Dipartimento di Fisica dell'Universit{\`a} di Pisa and INFN, Italia}\\
\end{center}

\bigskip

\centerline{\large\bf\color{blus} Abstract}
\begin{quote}\large
We study how present data probe standard and non-standard
properties of neutrinos and the possible existence of new light particles,
freely-streaming or interacting, among themselves or with neutrinos.
Our results include: $\sum m_\nu < 0.40\eV$ at $99.9\%$ C.L.;
that extra massless particles have abundance $\Delta N_\nu = 2\pm1$ if freely-streaming and
$\Delta N_\nu = 0 \pm 1.3$  if interacting; 
that 3 interacting neutrinos are disfavored at about $4\sigma$.
We investigate the robustness of our results by fitting different sub-sets of data.
We developed our own cosmological computational tools, somewhat different from the standard ones.
\color{black}
\end{quote}

\bigskip

\section{Introduction}
Thanks to recent data about  the Cosmic Microwave Background (CMB), Large Scale Structures (LSS) and also Type Ia Supernov\ae\ (SNe), cosmology has become the most sensitive probe of some neutrino properties (e.g.\ of neutrino masses: oscillation experiments test squared-mass 
differences, and other means of probing the absolute neutrino mass are currently less sensitive) and a  very sensitive probe of other neutrino properties, including non standard ones~\cite{review,LesgPasReview,BS}. In this paper we study how present cosmological data determine standard and non-standard `neutrino cosmology'. This includes three different issues.
\begin{itemize}
\item[i)] testing neutrinos: their masses, abundances,  \ldots
\item[ii)]  do photons, neutrinos and gravitons
make up the complete list of light particles? Data from particle physics allow extra light particles that are 
neutral under the Standard Model (SM) gauge group, and
such extra light particles appear in many speculative extensions of the SM,
one interesting example being simpler string models.\footnote{Within the string scenario
light particles can be avoided at the price of assuming that strings
vibrate on complicated enough  higher dimensional geographies~\cite{AntroString},
such that predictivity seems lost.}

\item[iii)] The two above issues can be connected, because neutrinos are the least tested light particles and can easily interact with new light neutral particles, in a way that affects the evolution of cosmological inhomegeneities. 

\end{itemize}
In section~\ref{th} we characterize the fundamental theories and
describe the cosmological parameters that we want to extract from present data.
Since our implementation of the cosmological computational tools needed for this analysis
somewhat differs from the standard one, we describe it in section~\ref{tool}.
Section~\ref{res} describes our results (table \ref{tab:navigator} might help in navigating the paper),
summarized in the conclusions.

\begin{table}[t]
\begin{center}
\begin{tabular}{r|lr|lr|lr|lr}
& \multicolumn{2}{c|}{Neutrino}   & \multicolumn{6}{c}{Cosmology with extra light particles}  \\ 
& \multicolumn{2}{c|}{cosmology} & \multicolumn{2}{c|}{freely-streaming} & \multicolumn{2}{c|}{self-interacting} & \multicolumn{2}{c}{interacting with $\nu$} \\
\hline
& & & & & & & & \\[-1.9mm]
massless &\parbox{8ex}{$A_s,n_s,h$,\\
$\Omega_b,\Omega_{\rm DM},\tau$} & \S\ref{tool}  & $\Delta N_\nu$ & \S\ref{sec:Nnu} & $\Delta N_\nu$ & \S\ref{NnuI} & $N_\nu,R\equiv{\displaystyle \frac{N_\nu^{\rm normal}}{N_\nu }}$ & \S\ref{00fluid} \\[8mm]
\ massive&$\sum m_\nu$ &\S\ref{sec:mNu}&$\Delta N_\nu, m_{\rm s}$ & \S\ref{sec:nus}&$\Delta N_\nu, m_{\rm s}$&\S\ref{sec:NnuIm}&$R=0, m_\nu$  or $m_\phi$ &\S\ref{m0fluid},\ref{0mfluid} \\
\end{tabular}
\caption{\label{tab:navigator}\em Schematization of the cases considered in this paper. For each one we list the notation of the relevant parameters probed by cosmology and refer to the relevant part of the text.}
\end{center}
\end{table}



\section{Theory}\label{th}
The effects of  new light particles  on the evolution of cosmological inhomogeneities
are often presented in terms of  ``non standard neutrino properties''  because 
i) particle physics has tested neutrinos less thoroughly than other particles, leaving room for surprises;
ii) neutrinos $\nu$ are the particles that can naturally interact with some new light neutral states.
In this section, we discuss how established data and theory restrict the behavior of  possible new light states, and in the rest of the paper we will consider scenarios that are compatible with these restrictions (more exotic possibilities are sometimes entertained in purely phenomenological analyses).

We consider light particles, neutral under all SM gauge interactions.
Indeed the LEP measurement of the invisible $Z$ width~\cite{Zwidth} implies that hypothetical new light particles
must be neutral under  $\SU(2)_L$ gauge interactions and can have at most a small hypercharge;
furthermore new light particles with strong interactions seem excluded.\footnote{On the experimental side, data can leave open windows.
On the theoretical side, light colored particles typically  form hadrons not lighter than the QCD scale;
one can however imagine  a colored scalar with a negative `tachionic' bare squared mass, fine tuned to almost cancel the QCD energy.
We name this speculative particle `rattazzon' in honor of the only theorist that accepted to discuss it.}
Field theory and gauge invariance significantly
restrict the interactions of new neutral light particles in a well-known way that depends on their spin.
\begin{itemize}
\item New light fermions, neutral under all SM gauge interactions (commonly called `sterile neutrinos' $\nu_{\rm s}$),
can have a mass mixing with ordinary `active' neutrinos. This is described by the following $\SU(2)_L\otimes{\rm U}(1)_Y$ gauge-invariant Lagrangian, written in compact notation
\beq \Lag(\nu_{\rm s}) 
= \Lag_{\rm SM} + \bar\nu_{\rm s}i \ds \nu_{\rm s} +(\lambda LH\nu_{\rm S} + \frac{M}{2}
\nu_{\rm s}^2 + \hbox{h.c.}).
\eeq
$L$ and $H$ are the lepton and Higgs weak doublets.
\item New light neutral scalars $\phi$ can have $\SU(2)_L\otimes{\rm U}(1)_Y$-invariant Yukawa interactions with {\em sterile} neutrinos, as
described by the following  Lagrangian:
\beq \Lag(\nu_{\rm s},\phi ) = \Lag(\nu_{\rm s}) +(\partial_\mu \phi)^2 + (\frac{\lambda'}{2}~\phi\, \nu_{\rm s}^2+\hbox{h.c.}) +
V(\phi).\eeq
At energies below the mass of a sterile neutrino, it can be integrated out
so that one effectively obtains Yukawa couplings $ \nu\nu \phi$, suppressed by the mass of the  sterile neutrino, as well as Yukawa couplings $ \nu\nu_{\rm s}\phi$ with other lighter sterile neutrinos.
We are aware of no reasonable allowed way of coupling a scalar directly to active neutrinos, 
in absence of light sterile neutrinos.

\item Sizable couplings of neutrinos to light vectors can be introduced in a similar way: 
adding sterile neutrino(s)  charged under an extra spontaneously broken gauge symmetry.
\end{itemize}
The gauge and Lorentz symmetries  allowed theorists 30 years ago
 to successfully predict and guess  experimental results.
Theorists have tried to proceed further by demanding a further restriction: naturalness.
At the moment it is unclear whether this is a correct requirement;
for instance, the LHC will tell whether physics at the weak scale obeys it or not.
Its imposition would strongly restrict the behavior of new light scalars.
For example, one could explain their lightness by assuming that they are Goldstone bosons
of spontaneously broken lepton numbers: in models with a single scalar, this implies that it couples
to neutrino mass eigenstates (rather than to generic combinations), forbidding neutrino decay in vacuum.

\bigskip

Even taking into account that these considerations significantly restrict the
set of  `reasonable'  fundamental theories, 
flavour issues make the number of fundamental parameters so large that a direct study
of the fundamental theory seems unpractical.
(This approach has been pursued in~\cite{CMSV} in the case of a single sterile neutrino and no bosons).
On the other hand, cosmology probes (new) light particles via their gravitational couplings:
neutrinos and new light particles affect the evolution of inhomogeneities in the gravitational
potential(s), in which matter moves.
Therefore it is convenient to focus on parameters  half-way between theory and data:  equation of state, sound speed, etc. 
We will consider the following basic limiting cases: extra freely-streaming particles, or
extra particles that interact among themselves, or extra particles that interact with neutrinos.
Whether a particular model falls in one or the other category depends  on the value of the fundamental couplings (like $\lambda$, $\lambda'$ and the interactions in $V(\phi)$ above) of the particular model.

\medskip

In the following analyses we will study different systems specifying the parameters probed by  cosmology. Some of these parameters will be combinations of the parameters appearing in
the low-energy effective Lagrangian, some others will not (e.g.\ the abundances around recombination).
The latters could be translated in terms of the fundamental parameters of each full model, 
a step that we do not make. 

We ignore other possible probes of the same physics
(which e.g.\ also manifests as
neutrino decay, anomalous matter effects, rare $K\to \ell\nu$ decays), as they presently
are orders of magnitude  less sensitive than cosmology.

\begin{table}[t]
\begin{center}
\begin{tabular}{c|cc}
& `Typical' analyses & This analysis \\ \hline 
\\
Cosmological code &public CMBfast & our code \\
Language & \parbox{5cm}{\begin{center}  {\footnotesize\textsc{FORTRAN}}\end{center}} & \parbox{6cm}{\begin{center}Mathematica\end{center}} \\
Statistics & Monte Carlo Markov chains & Gaussian\\
\end{tabular}
\caption{\label{tab:dif}\em Differences between our approach and a `typical' cosmological analysis.}
\end{center}
\end{table}

\begin{figure}[t]
$$\includegraphics[width=0.95\textwidth]{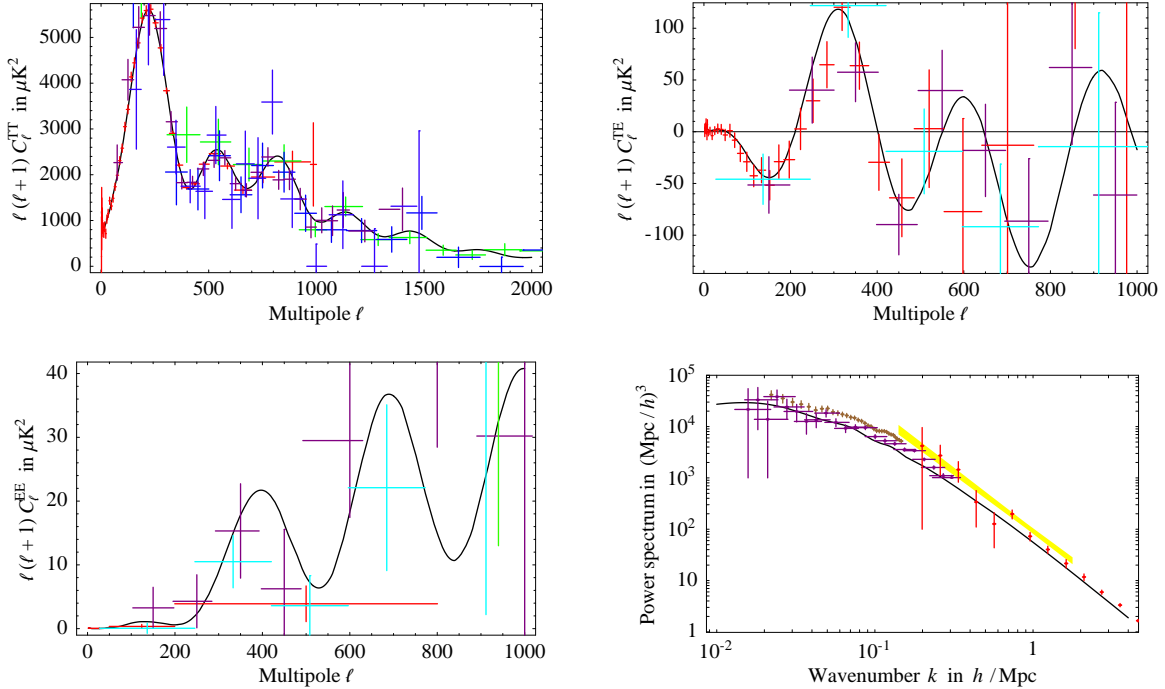}$$
\caption{\label{fig:data}\em Our computation of CMB and LSS spectra in standard {\rm $\Lambda$CDM} cosmology, compared with data.}
\end{figure}

\begin{figure}[t]
$$\includegraphics[width=0.95\textwidth]{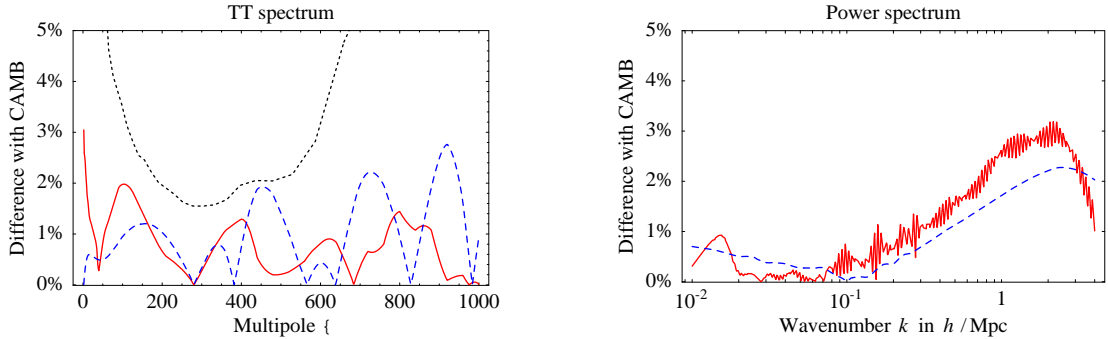}$$
\caption{\label{fig:diff}\em Difference between our code and CAMB, at 
the standard-cosmology  best-fit point for $m_\nu=0$ (red solid line) and for $m_\nu=0.5\eV$ (blue dashed line).
Our code does not employ any approximation specific for standard cosmology.
In both codes various parameters allow the user to increase the accuracy; 
this plot holds for the choice employed in the present paper.
The dotted line shows the $1\sigma$ accuracy obtained by WMAP3 results (binned data),
indicating that we have a good enough accuracy (as confirmed by other tests).
A similar $\%$-level accuracy is found for the {\rm TE} and {\rm EE} CMB spectra, 
that presently are measured with much larger uncertainties than the {\rm TT} spectrum.}
\end{figure}

\begin{figure}[t]
$$
\includegraphics[width=0.95\textwidth]{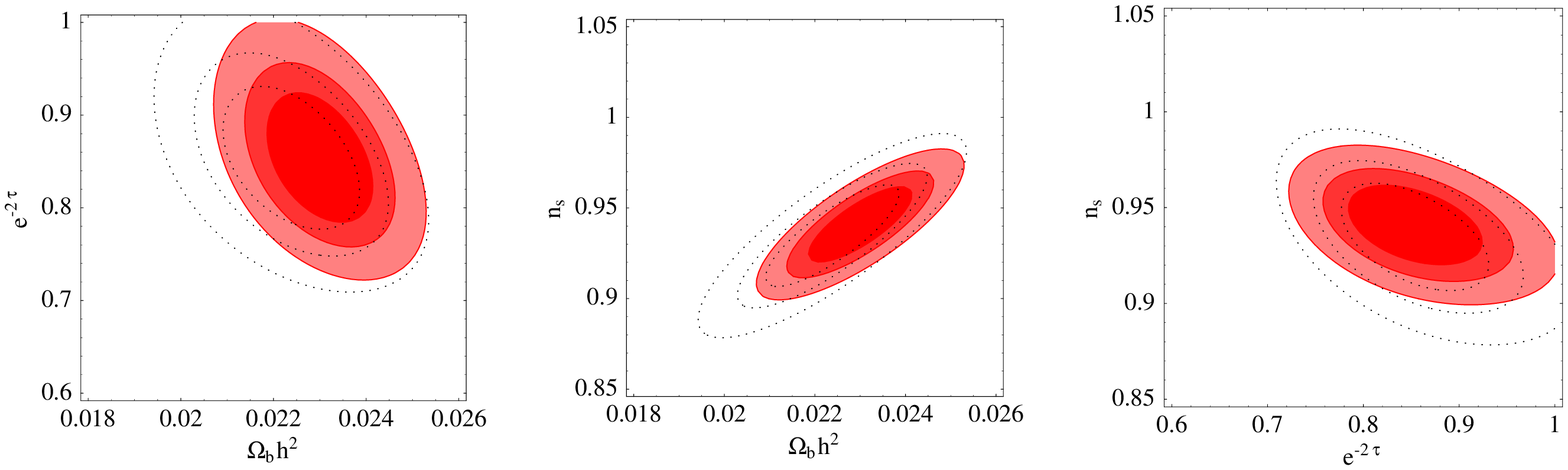}$$
\caption{\label{fig:StandardCosmo}\em Fit of cosmological data at 68, 90 and 99\% C.L.
The shaded areas show our global fit without Lyman-$\alpha$, 
and the dotted lines our WMAP3-only fit, such that this figure can be directly
compared with the analogous WMAP Science Team plots in fig.\ 10 of~\cite{WMAPparam}.
}
\end{figure}

\section{Analysis strategy}\label{tool}
We develop our own computational tools for the analysis of the cosmological observables.
For what concerns standard cosmology, our results agree with those of other authors (e.g.\ the WMAP  Science Team), but having independent analyses is clearly important.
In this respect, our analysis is particularly independent: it 
differs from what nowadays is a typical analysis in the way illustrated in table~\ref{tab:dif}.
Cosmological observables are computed using a code written by one of us, rather than
running the commonly used CMBfast or CAMB public codes~\cite{CMBfast}:
this allows us to have a better control and flexibility on non-standard modifications.

We use the line-of-sight approach in the conformal Newtonian gauge~\cite{CMBfast, Seljak94, MB};
recombination can be implemented both in Peebles approximation (see e.g.~\cite{Dodelson}) 
and using the external {\tt recfast} code~\cite{recomb}, which is the option chosen for the present analysis.

The main disadvantage is that our code is almost $2$ orders of magnitude slower than CMBfast or CAMB.
In part this happens because, rather than optimizing our code for standard cosmology,
we keep it fully flexible such that non-standard cosmologies are immediately implemented.\footnote{For example,
the interacting particles considered in this paper are implemented by typing their linear evolution equations, eq.\eq{fluidevo0} or eq.\eq{fluidevo},
in {\tt NDSolve} form.}
In part this happens because, while standard codes are written 
in {\footnotesize\textsc{FORTRAN}}, our code is written in Mathematica~\cite{Mathematica} and we run it on a common laptop (rather than on a cluster of computers).

We now describe the advantages of our approach that allowed us to perform our analysis.
Readers not interested in these technical details can skip the rest of this section.
The main point is that, while {\footnotesize\textsc{FORTRAN}} can only do numerical computations,
Mathematica does not have this limitation and  allows to do analytically all parts
of the computations that can be done analytically.
This includes the dependence of cosmological observables, e.g. on the spectral index, 
and all statistical issues that nowadays are the most time-consuming aspect
of cosmological analyses.
Our approach is based on the powerful old-fashioned Gaussian techniques, as we now briefly describe.

\subsection{Statistics}\label{Statistics}
Cosmological data have become so accurate and rich that
debates about Bayesian priors versus frequentistic constructions are getting numerically irrelevant:
all different techniques converge towards their common gaussian limit.
This is clear e.g.\ from figures~10 of the WMAP analysis~\cite{WMAPparam}:
within good approximation all allowed regions identified
by their Monte Carlo Markov Chain (MCMC) technique are ellipses
(with sizes that have the Gaussian dependence on the confidence level),
as they must be in Gaussian approximation.
This means that the usual $\chi^2$,
a single quadratic function of the various cosmological parameters,
approximatively encodes all present information on standard cosmology and that the dependence on the $N_p^{\rm stnd}$ parameters of standard cosmology
(here chosen to be the usual $A_s,n_s,\Omega_{\rm DM},\Omega_{\rm b},Y_p,h,\tau$
with $\Omega_{\rm tot}=1$, defined as in~\cite{WMAPparam}; $A_s$ is normalized at the pivot
point $k=0.002/{\rm Mpc}$) is accurately enough described by a first order Taylor expansion
of each observable (the various $C_\ell^{TT}$, $C_\ell^{TE}$, $C_\ell^{EE}$, the power spectra,
the luminosity distances of supernov\ae, ...) around any point close enough to the best-fit point.
We will soon check explicitly that sampling $N_p^{\rm stnd}+1\circa{<}10$ points is enough to study standard cosmology.\footnote{We do not improve the accuracy by making a second-order Taylor expansion. This can be done by probing $N_p^{\rm stnd}(N_p^{\rm stnd}+1)/2\circa{<} 50$ more points only,
but would complicate statistical issues, preventing e.g.\ analytical marginalizations
of the likelihood over nuisance parameters.

Furthermore, we checked that using two-sided derivatives or recomputing observables with the public CAMB code~\cite{CMBfast} affects the results of the global standard fit, eq.\eq{means}, in a minor way:
best-fit values typically shift by about a third of a standard deviation. 
}
For comparison, MCMC techniques can need one independent 
chain with $\sim 10^5$ points every time one wants to analyze a (sub)set of data~\cite{WMAPparam}.

The Gaussian approximation has no `statistical' uncertainty due to finite MCMC sampling but it introduces a `systematic' uncertainty.
This is small near the expansion point (chosen to be close to the best-fit point) and grows when one goes far from it.
At some point data become accurate enough that the region singled out by them is small enough to make the Gaussian approximation a good one.
By construction, the Gaussian approximation reproduces the same best-fit point
(small differences between our and other analyses on common studies are due to different data-sets, different code, etc.) and the confidence regions with
small enough confidence levels, and fails at larger confidence levels.
In practice, we care about $90\%$, $99\%$ and maybe $99.9\%$ confidence levels. Fig.\fig{StandardCosmo} is our crucial test and it shows that the contours corresponding to such confidence levels are reproduced in an fairly accurate way when comparing with the WMAP Science Team analysis.
Notice that the Gaussian approximation needs not to be and is not accurate enough to analyze every single piece of data, but it allows to correctly fit the full data-set.
Some non-standard cosmological parameters are still subject to `degeneracies':
we will later discuss how the  Gaussian approximation can be extended to  deal with these situations.

Our code directly gives the $\chi^2$ as an analytic quadratic function of cosmological parameters,
that fully describes present information on $\Lambda$CDM cosmology.
Our result in terms of best fit points and 1$\sigma$ errors is{\small 
\beq\label{eq:means}\begin{array}{ccccccc}
\hbox{fit}& A_s & h & n_s & \tau &100\Omega_b h^2 & \Omega_{\rm DM} h^2 \\
\hbox{WMAP3} &
0.80 \pm 0.05  &  0.704 \pm 0.033  &  0.935 \pm 0.019  &  0.081 \pm 0.030  &  \
2.24 \pm 0.10  &  0.113 \pm 0.010\\
\hbox{Global} &
0.84 \pm 0.04  &  0.729 \pm 0.013  &  0.951 \pm 0.012  &  0.121 \pm 0.025  &  \
2.36 \pm 0.07  &  0.117 \pm 0.003\\
\end{array}\eeq}
The symmetric correlation matrix for the global fit is\footnote{
This is closely related to the Fisher matrix widely used by cosmologists.
Since different communities use different terminologies, we here define the precise
meaning of the quantities we employ.
The mean values $\mu_i$ of the parameters $p_i$,
their errors $\sigma_i$ and the correlation matrix
$\rho_{ij}$ determine the $\chi^2$ as
\beq\label{eq:chiq}
\chi^2 =\sum_{i,j} (p_i - \mu_i) [(\sigma^2)^{-1}]_{ij}  (p_j - \mu_j),\qquad\hbox{where}
\qquad (\sigma^2)_{ij} = \sigma_i \rho_{ij} \sigma_j\  . \eeq
The likelihood is given by ${\cal L} = {\cal L}_{\rm max} e^{-\chi^2/2}$.
In Gaussian approximation, marginalizing  ${\cal L}$ with respect to
any sub-set of `nuisance' parameters is  equivalent to minimizing the
$\chi^2$ with respect to nuisance parameters.
E.g., since eq.\eq{chiq} contains the inverse of the $\sigma^2$ matrix, 
the $\chi^2$ marginalized
over all parameters $p$ except one of them, $p_i$, is
$\chi^2(p_i)=(p_i-\mu_i)^2/\sigma_i^2$.
}
\beq\label{eq:corr}\rho=
\left(
\begin{array}{llllll}
 1 & - & - & - & - & - \\
 -0.19 & 1 & - & - & - & - \\
 -0.25 & 0.71 & 1 & - & - & - \\
 0.74 & 0.31 & 0.42 & 1 & - & - \\
 -0.08 & 0.70 & 0.78 & 0.41 & 1 & - \\
 0.07 & -0.03 & 0.14 & -0.02 & 0.36 & 1
\end{array}
\right).
\eeq
Fig.\fig{data} shows our best-fit, and fig.\fig{diff} shows the percentage difference
in the main observables among our code and CAMB~\cite{CMBfast}.

\subsection{The data set}\label{dataset}
Our global fit includes the following set of data: 
\begin{itemize}
\item WMAP: the 3 year TT, TE and EE data from WMAP~\cite{WMAPdata} (called WMAP3 throughout).
We use data in binned form (72 bins) and deduce the Gaussian approximation of the full likelihood 
by extracting mean values, errors and their correlations 
from the numerical likelihood code provided by the
WMAP collaboration~\cite{Lcode}, that includes the uncertainty due to cosmic variance.
At small $\ell$ there is a non-Gaussian large uncertainty; a fortiori these data do not have a dominant statistical weight.
We verified that using data in unbinned form makes a very minor difference.

\item  Other CMB data  (103 bins): 
the most recent data from  ACBAR, {\sc Boomerang} (TT, TE and EE results), CAPMAP (EE), CBI (TT and EE), DASI (TE and EE), VSA~\cite{CMB}.

\item Large Scale Structures: 
\begin{itemize}
\item The SDSS~\cite{SDSS} and 2dF~\cite{2dF} measurements of the matter power spectrum from galaxy surveys (called LSS throughout). An important and long-standing issue in this respect is how to relate the quantity of interest for cosmology (the matter power spectrum $P(k)$, computed with linear perturbation theory) to the measured galaxy-galaxy power spectrum $P_{\rm gal}(k)$. We adopt the prescription suggested in~\cite{2dF} to effectively take into account the normalization bias and the nonlinear effects: $P_{\rm gal}(k) = b^2 (1+Q k^2)/(1+A k)P(k)$. We adopt $A=1.4$ and we marginalize over a free bias parameter $b$ and over $Q = 4.6 \pm 1.5$ and $10 \pm 5$ for 2dF and SDSS respectively (we assume Gaussian errors), following~\cite{2dF} and~\cite{Seljak06}.

\item The detection of Baryon Acoustic Oscillation peaks (BAO) in the correlation function of the SDSS subsample of Luminous Red Galaxies at nominal redshift $z=0.35$~\cite{BAO}. We implement it in terms of a measurement of the adimensional parameter 
$$A=D_{\rm v}(0.35) \sqrt{\Omega_{\rm matter} H_0^2}/(0.35\ c) = 0.469 \pm 0.017$$
(neglecting  a small dependance on the primordial spectral index $n_s$), where $c$ is the speed of light and $D_{\rm v}$ is a distance defined in terms of the comoving angular diameter distance $D_{\rm A}$ and the Hubble parameter $H$ as $D_{\rm v}(z)=\left[D_{\rm A}(z)^2 c z/H(z)\right]^{1/3}$.
We computed the galaxy-galaxy correlation function in a few cases, checking that this is a satisfactory approximation.
Within standard cosmology, BAO data give a measurement of the total matter density:
\beq
\Omega_{\rm matter} = 0.28\pm0.025~\cite{BAO}.\eeq

\item The Lyman-$\alpha$ Forests in distant quasars absorption spectra (called Lyman-$\alpha$ throughout) from Croft et al.~\cite{LyCroft}, at redshift $z=2.72$, and from SDSS as condensed by~\cite{LySDSS} into the measurement of the renormalized amplitude of the power spectrum $\Delta^2(k,z) = k^3 P(k,z)/(2 \pi^2) = 0.452 \pm 0.072$ and of its slope $n_{\rm eff}(k,z) = d \ln P(k)/ d \ln k = -2.321 \pm 0.069$ at the pivot point $k=0.9\ h/{\rm Mpc}$ and $z=3.0$.\footnote{We reduced the errors by $25\%$ 
because this brings the Gaussian $\chi^2$ in closer
agreement with the `exact' likelihood as provided in~\cite{LySDSS}.} Different re-analyses of the same data find somewhat different results (see e.g.~\cite{LyViel,Seljak06}), which might signal systematical problems, e.g. in the way the flux power spectra are converted into measurements of the matter power spectra. In the following we will adhere to the values presented in the data sets above, but pay special attention to the implications of their use.
\end{itemize}

\item Type I supernov\ae: the Gold sample of Riess et al.~\cite{SNe1} and the SNLS05 data~\cite{SNe2}.
We combine the datasets directly, although strictly speaking they are not independent as they share the same set of low-redshift supernov\ae\ and use slightly different techniques in the analysis;  
however these simplifications marginally affect the final results
such that a fully careful combination seems unnecessary; present cosmological data surely contain more worrisome issues.
Within standard cosmology, supernov\ae{} data give measurements of the total matter density,
\beq
\Omega_{\rm matter} = 0.33\pm0.036~\cite{SNe1},\qquad
\Omega_{\rm matter}=0.26\pm0.038~\cite{SNe2}.\eeq
\item The HST measurement of the Hubble constant $H_0 = 100\,h\,{\rm km/sec/Mpc}$:
$h=0.72\pm0.08$~\cite{HST}.

\end{itemize}
All CMB data are marginalized with respect to the recombination and
Sunyaev-Zeldovich backgrounds.

\bigskip

Our strategy concerning statistics needs to be extended to study
the new physics scenarios addressed in the next section.
In general the dependence of the $\chi^2$ on the new physics parameters is not Gaussian,
because we consider small new physics effects that behave in significantly different
ways in different regions of their parameter space.

In most cases (e.g.\ massive neutrinos, section~\ref{sec:mNu}) the simplest extension
is enough: we compute the observables as function of the new physics parameters 
(e.g.\ for a dozen of values of neutrino masses) and, at each point, we use the Gaussian technique to marginalize over
the standard cosmological parameters listed in eq.\eq{means}.

In other cases (e.g.\ in presence of extra massless neutrinos, section~\ref{sec:Nnu})
this turns out to be not enough, because data still allow for large new physics effects:
e.g.\  a large variation in $N_\nu$ can be
compensated,  to a large extent, by readjusting the other cosmological parameters, expecially
$h$ and $\Omega_{\rm DM}$~\cite{melchiorri}.
We deal with this kind of `degeneracy' by applying the
well-known Newton iterative minimization technique:
for any given value of the new parameters (e.g.\ $N_\nu$) 
we perform the Gaussian expansion around some point:
in general it gives an inaccurate value of the minimum $\chi^2$ because the expansion
point is not close enough  to the best-fit point:
we then use the just estimated best-fit point to perform a second Gaussian expansion around it,
finding an improved approximation to the minimum $\chi^2$ and to the true best-fit point.
In practice, two or three iterations of this procedure are enough to find a good approximation,
as can be confirmed by checking that more iterations no longer change the result.
Subsets of data can then be analyzed by performing the Gaussian approximation around this global best-fit point.



\begin{figure}[t]
$$
\includegraphics[width=0.4\textwidth]{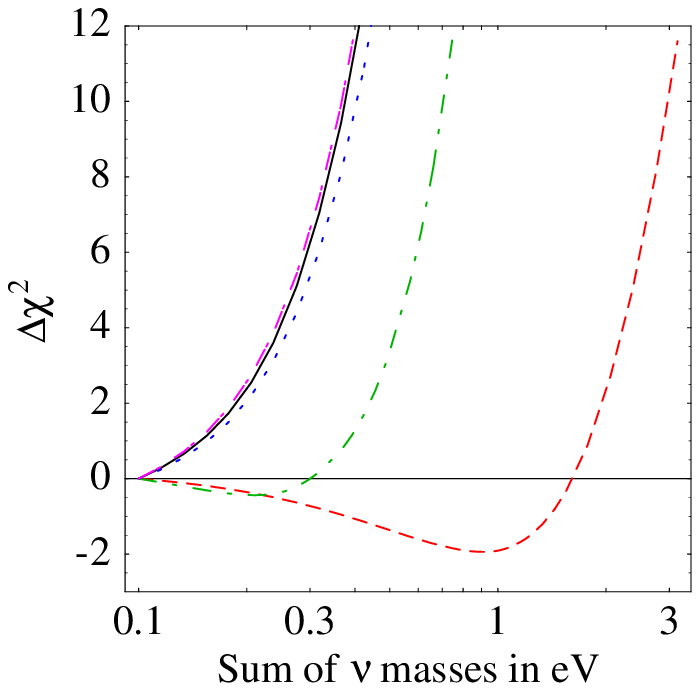}\qquad
\raisebox{0.02\textwidth}{\includegraphics[height=0.37\textwidth]{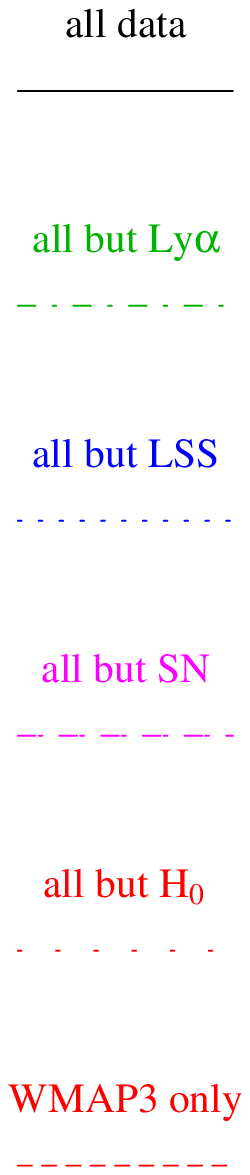}}$$
\caption[X]{\label{fig:Nm}\em 
Fit as function of the energy density in freely-streaming slow particles,
parameterized by the sum of active neutrino masses,
assumed to have the standard abundance $N_\nu=3.04$.
We studied different combinations of data-sets, as indicated by the legend.}
\end{figure}

\section{Results}\label{res}
In all cases we will report the
difference $\Delta \chi^2$ with respect to the standard $\Lambda$CDM model (which is individuated by the best fit points in eq.\eq{means} and assumes the existence of only 3 ordinary, massless neutrinos), because
this is the significant statistical indicator.
Roughly speaking, $\Delta\chi^2 +\Delta N_p= \pm n^2$ corresponds to 
$n$ standard deviations of evidence against or for,
where $ \Delta N_p = N_p-N_p^{\rm stnd} $ is the number of extra relevant free parameters
with respect to $\Lambda$CDM.
For the models that we will consider, 
the not univocally defined $\Delta N_p\sim 1$ or 2 is small enough
that it does not play a significant r\^ole.
In practice, this means e.g.\ that $3\sigma$ ranges can be read from fig.\fig{Nm} or fig.\fig{NNu} by
looking where our curves reach $\Delta\chi^2 = 9$.

We here do not judge the compatibility of a model with data by performing the Pearson $\chi^2$ test.
According to this test, a model is excluded if the total $\chi^2$ is not compatible with its expected value
$N \pm \sqrt{2N}$ where $N$ is the number of data points minus the number of free parameters
(for $N\gg 1$ the $\chi^2$ follows a Gaussian distribution with this mean and variance;
for small $N$ the $\chi^2$ has a different `$\chi^2$' probability distribution).
The $\sqrt{2N}$ means that the Pearson $\chi^2$ test becomes inefficient when $N\gg 1$, 
which is the case in cosmological fits, where $N\sim (100\div 1000)$.
In practice, this means that a model disfavored at $5\sigma$ by the $\Delta \chi^2$ test
typically is disfavored only at $1\sigma$ by the $\chi^2$ test.
A more detailed 
discussion of these issues can be found in section 4 of~\cite{Pearson} 
(in the context of fits of solar neutrino data).



\subsection{Massive neutrinos}\label{sec:mNu}
Fig.\fig{Nm} shows the result of our comparison of cosmological data with massive neutrinos.  
We here assume that only the three active neutrinos exist, and have the standard abundance and temperature.
Cosmology is not yet sensitive enough to discriminate neutrinos with normal from inverted hierarchy,
but only to heavy quasi-degenerate neutrinos.
Therefore the key parameter probed by cosmology is the sum of neutrino masses
$\sum m_\nu = m_1+m_2 + m_3$.
Oscillation data imply $\sum m_\nu \approx 0.05\eV$ ($0.10\eV$)~\cite{review} if neutrinos have strictly normal (inverted)
mass hierarchy, and larger values of $\sum m_\nu$ are obtained for quasi-degenerate neutrinos.
We assumed normal hierarchy at $\sum m_\nu =0.05\eV$,
inverted hierarchy at $\sum m_\nu =0.1\eV$,
and degenerate neutrinos at larger $\sum m_\nu$.

\medskip

Within standard cosmology,
WMAP3 data alone imply a constraint not plagued by potential systematical problems:
we find $\sum m_\nu < 2.2\eV$ at $95\%$ C.L.\ (see fig.\fig{Nm})  in agreement with~\cite{WMAPparam,Fukugita}. 
Adding LSS data (that are more strongly affected by neutrino masses than CMB data)
gives stronger constraints, but they are subject to the potential systematic problems 
discussed in section~\ref{dataset}.
We therefore plotted different lines, that correspond to different combinations of data-sets,
with each kind of observable excluded in turn.
Our global fit gives
\beq \label{eq:mnubound}
\sum m_\nu < 0.40\eV \qquad\hbox{at 99.9\% C.L.}\eeq
in good agreement with~\cite{HannestadBAO} and in reasonably good agreement with~\cite{Seljak06}.
We see that SN data have essentially no effect, while Lyman-$\alpha$ data play an important r\^ole:
the constraint becomes about 2 times weaker if Lyman-$\alpha$ data are dropped:
$\sum m_\nu < 0.73\  (0.52) \eV$ at 99.9\% (95\%) C.L., in good agreement with~\cite{HannestadBAO}
and in reasonably good agreement with~\cite{WMAPparam}.

However, the constraint from the global fit is somewhat stronger than the sensitivity of the data.
Indeed, focussing on Lyman-$\alpha$ data, 
the main effect of neutrino masses at fixed values of other cosmological parameters, 
consists in reducing the adimensional $\Delta^2$, measured by SDSS Lyman-$\alpha$ data 
with $\pm 0.07$ uncertainty, by
about $0.13 \sum m_\nu/\eV$.
However, for massless neutrinos, the global fit already suggests a value of $\Delta^2$ about 2 standard deviations below
the experimental value, and making neutrinos massive reduces further the expected $\Delta^2$.
In eq.\eq{mnubound} we  avoided reporting a strong but doubtful constraint by
choosing a confidence level higher enough than 2 standard deviations. 


\subsection{Extra freely-streaming massless  particles}\label{sec:Nnu}
\begin{figure}[t]
$$
\includegraphics[width=0.4\textwidth]{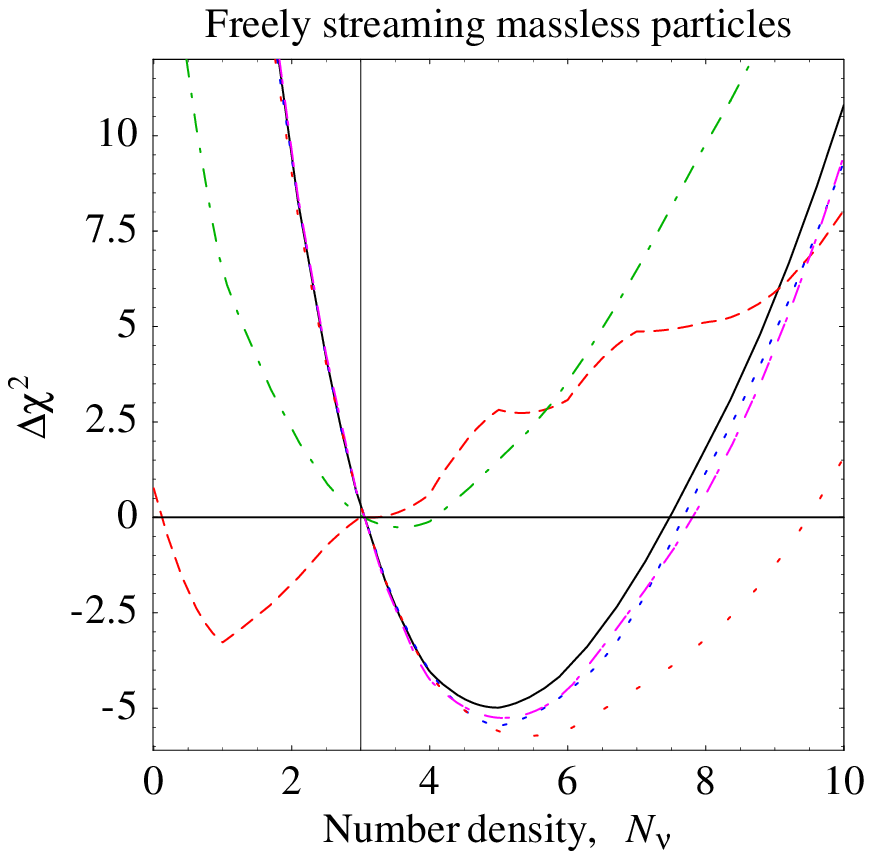}\qquad
\raisebox{0.02\textwidth}{\includegraphics[height=0.37\textwidth]{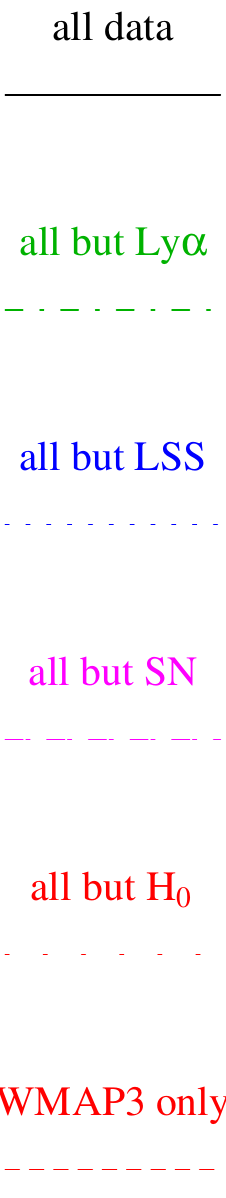}}\qquad
\includegraphics[width=0.4\textwidth]{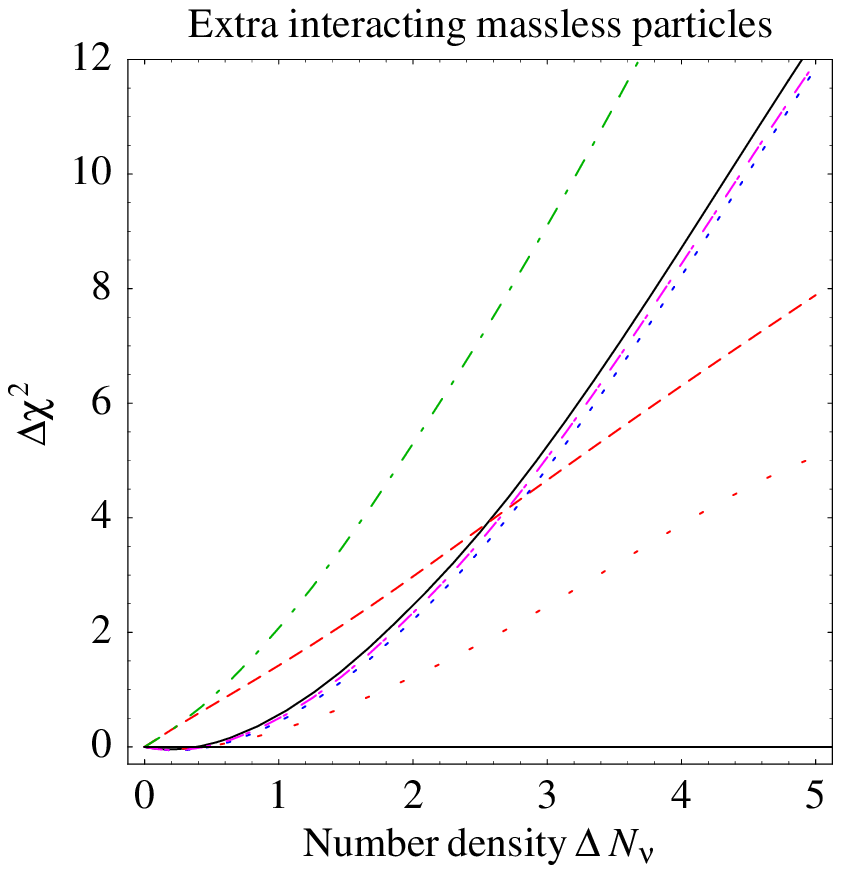}$$
\caption[X]{\label{fig:NNu}\em 
Fig.\fig{NNu}a): fit as function of the energy density in freely-streaming relativistic particles,
parametrized by the usual `number of neutrinos' $N_\nu$.
Fig.\fig{NNu}b): fit as function of the energy density in extra interacting relativistic particles,
with abundance parametrized by $\Delta N_\nu$.
We studied different combinations of data-sets, as indicated by the legend.
}
\end{figure}

As usual the `number of neutrinos'  $N_\nu$ parameterizes the amount of energy in all relativistic freely-streaming degrees of freedom, converted in terms of `neutrino equivalents': $N_\nu$ includes the ordinary neutrinos and any extra fermion or boson, as it is formally defined by the relation $\rho_{\rm relativistic} = \rho_\gamma\left[1 + 7/8\ N_\nu (T_\nu/T)^4 \right]$, where $T_\nu/T = (4/11)^{1/3}$
at $T\ll m_e$.
Standard cosmology with 3 neutrinos predicts $N_\nu \approx 3.04$ (the deviation from 3 being due to the incomplete $\nu_e$ decoupling at the time of the beginning of $e^+e^-$ annihilations, plus other small corrections).
Our global fit gives
\beq N_\nu =5 \pm 1\eeq
and  the continuous line in fig.\fig{NNu}a shows the precise form of the $\Delta \chi^2$.
We checked that using
the first release of WMAP data our result agrees with the various analyses published in the literature~\cite{Nnu1}.

In general, results of global fits can be mislead by problems in any of the pieces of data
they contain; in this case the validity of the global fit appears particularly doubtful:
$N_\nu$ is dominantly determined by non-CMB data, and
giving slightly different weight to them can significantly affect the fit because
different pieces of data prefer different values of $N_\nu$.
In particular, the $2\sigma$ preference for $N_\nu>3$ is mainly due to the $2\sigma$ anomaly
in the Lyman-$\alpha$ measurement of the power spectrum: 
fig.\fig{NNu}a shows that omitting Lyman-$\alpha$ one recovers excellent agreement with the standard value $N_\nu =3$.
The agreement with and between the up-to-date analyses performed by the WMAP Team~\cite{WMAPparam} and by~\cite{Seljak06} is imperfect; in particular  the revised version of~\cite{Seljak06} claims a $3\sigma$ preference for $N_\nu >3$.



\subsection{Extra massless particles interacting  among themselves}\label{NnuI}
In the previous section we considered extra (massless) particles with negligible interactions,
that therefore freely move on cosmological scales.
We now consider the opposite limit:  extra (massless) particles that interact among themselves
with a mean free path smaller
than relevant cosmological scales, such that inhomogeneities in their energy density 
evolve in a different way.
Concrete examples are an elementary scalar with a quartic self-interaction
or any particle with low compositeness scale, 
obtained e.g.\ if some extra QCD-like gauge group becomes
strongly coupled at an energy much lower than the QCD scale.
In the tight coupling limit this system is described by a fluid: 
its density and velocity perturbations $\delta$ and $v$ obey the standard fluid equations
(in the conformal Newtonian gauge and in linear approximation):
\beq\label{eq:fluidevo0} \dot\delta = - 4 \dot\Phi-\frac{4}{3} k v ,\qquad
\dot v = k\Psi + \frac{ k\delta}{4}
\eeq
where a dot denotes derivative with respect to conformal time,
$k$ is the wavenumber, $\Phi$ and $\Psi$ are the scalar perturbations in the metric
(in the notations of~\cite{Dodelson}).

Fig.\fig{NNu}b shows the constraint on the density of the extra particles, that we parameterize
in terms of the usual `equivalent number of neutrinos' $\Delta N_\nu\ge 0$:
the global fit gives
\beq
\Delta N_\nu =0\pm1.3.
\eeq
Data do not favor the presence of extra massless particles interacting  among themselves.
Dropping Lyman-$\alpha$ data makes the $1\sigma$ constraint two times more stringent.
Fig.\fig{NNu}b shows that
the interval allowed at $n$-$\sigma$ is not $n$ times larger than the $1\sigma$ interval.

\begin{figure}[t]
$$
\includegraphics[width=0.45\textwidth]{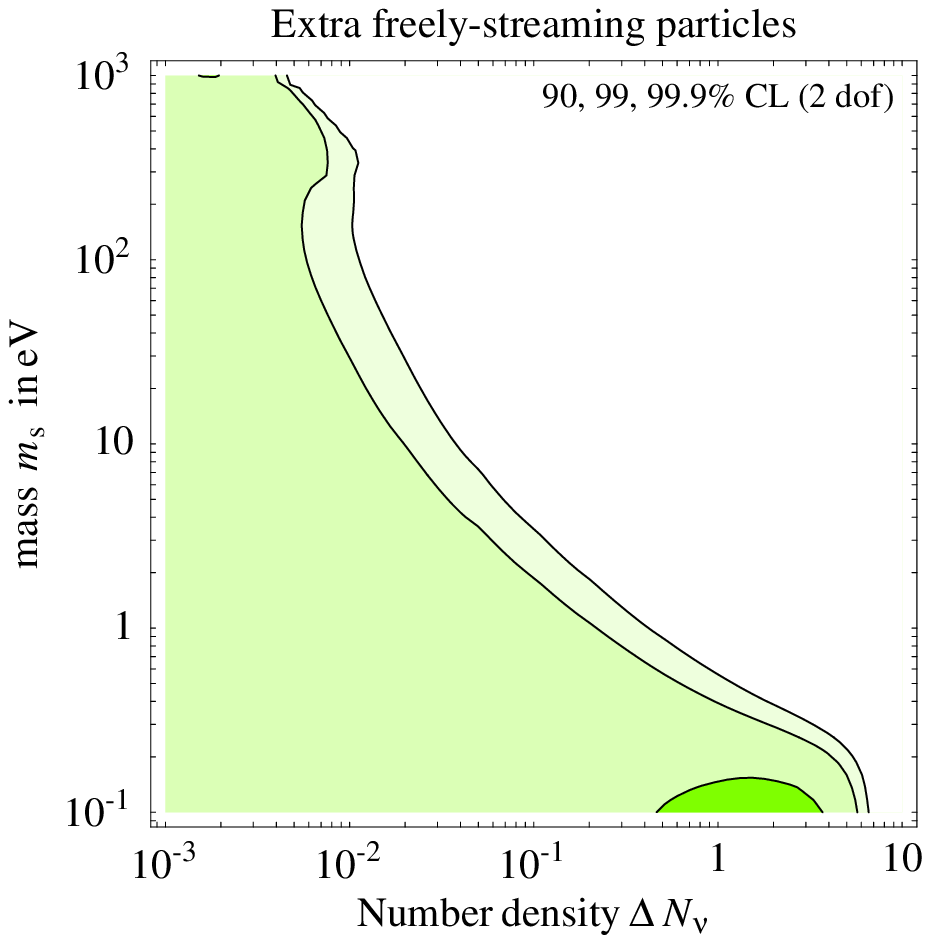}\qquad
\includegraphics[width=0.45\textwidth]{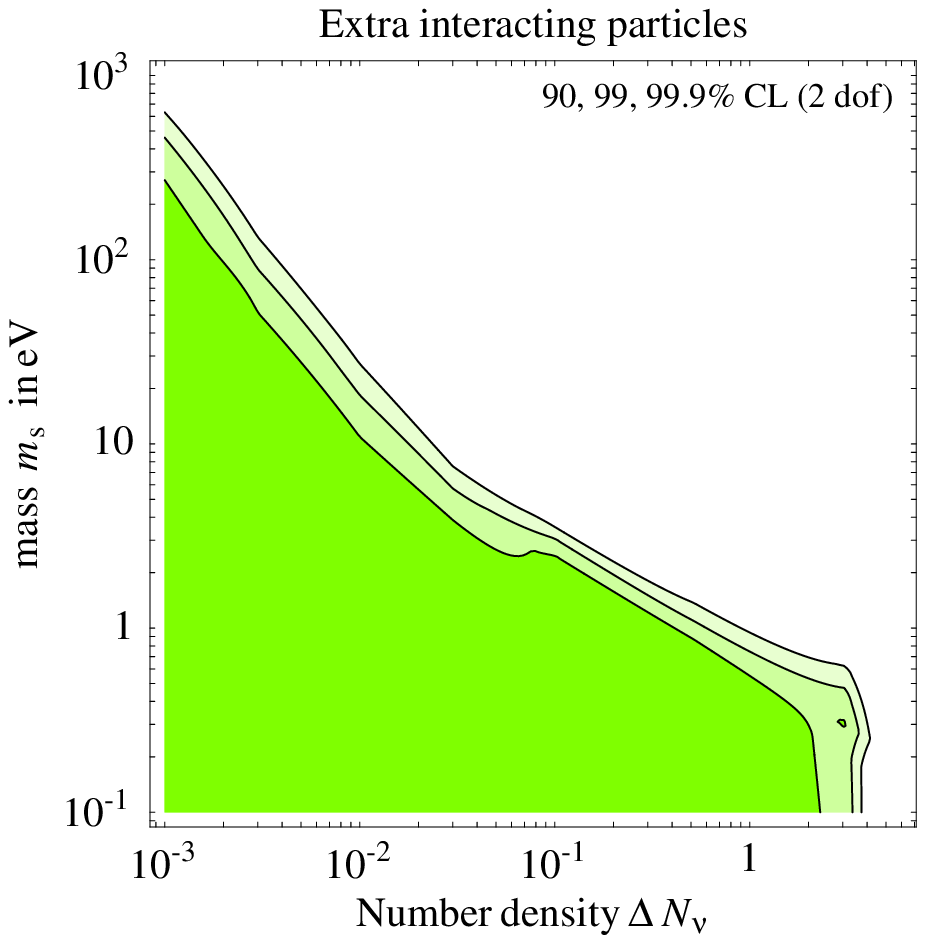}$$
\caption[X]{\label{fig:mN}\em Fig.\fig{mN}a: how the global fit constrains the presence of
extra freely-streaming particles with abundance $\Delta N_\nu$ and mass
$m_{\rm s}$.  Shaded regions are allowed.
Fig.\fig{mN}b: the analogous result for extra interacting particles.
}
\end{figure}

\subsection{Extra freely-streaming massive  particles}\label{sec:nus}
We here assume that the ordinary neutrinos have cosmologically negligible masses (e.g.\ $0.05\eV$~\cite{review}), 
and add extra freely-streaming particles, with abundance $\Delta N_\nu$, mass $m_{\rm s}$ and the same temperature as the ordinary neutrinos.
We have chosen the notation $\Delta N_\nu$ and $m_{\rm s}$ because,
although the constraints we obtain here apply to a generic freely-streaming fluid (but not to extra interacting particles, studied in the next sections), 
sterile neutrinos of mass $m_{\rm s}$ are by far the most popular specific realization.
Sterile neutrino models allow one to compute the abundance $\Delta N_\nu$ in terms of
oscillation parameters  (non-minimal scenarios introduce extra parameters, such as particle/anti-particle asymmetries), 
and $\Delta N_\nu$ can acquire a small, sub-thermal value, that
stays constant after neutrino decoupling at $T\sim {\rm MeV}$.
We here assume that $\Delta N_\nu$ keeps an arbitrary constant value at the much lower temperatures relevant for CMB and LSS observables.
The present energy density is
\beq  \Omega_{\rm s} \equiv \frac{\rho_{\rm s}}{\rho_{\rm cr}}
= \frac{0.01}{h^2}\frac{m_{\rm s}\cdot \Delta N_\nu}{\eV}.\eeq

This case is characterized by 2 new parameters: therefore we must make 2-dimensional plots, so we prefer not to show how the fit changes by considering 
various different data-sets.
Fig.\fig{mN}a shows the global fit.
We can distinguish three regions.
For small $m_{\rm s}$ and large $\Delta N_\nu\circa{>} 1$, data disfavor a too large number of
relativistic particles around recombination.
In the intermediate region the extra particles behave as warm Dark Matter:
their abundance is constrained to be  $\Omega_{\rm s}\circa{<}0.01$ dominantly by
Lyman-$\alpha$ and LSS data.
For larger $m_{\rm s}\gg \,{\rm keV}$ the extra particle
behave as cold enough Dark Matter and sterile densities 
as large as $\Omega_{\rm s}\approx 0.22$ are allowed.\footnote{Recent  dedicated analyses found that
the case in which sterile neutrinos provide all the observed Dark Matter, $\Omega_{\rm DM}\approx 0.22$, and (as we assume) have the same temperature of active neutrinos, 
is allowed by cosmological data only for $m_{\rm s}\circa{>}10\,{\rm keV}$~\cite{sterileDM}. (Such large masses are incompatible with X-ray observations, so that this scenario is allowed only if these particles are produced with sub-thermal velocities).
We limit our plots to the region $m_{\rm s}< 1\,{\rm keV}$, where the allowed
$\Omega_{\rm s}$ is small enough that our Gaussian technique (implemented in the most na\"{\i}ve way)
provides a good approximation.  
The structure present in our plot at $m_{\rm s}\circa{<} {\rm keV}$
might be artificially too sharp, because Lyman-$\alpha$ SDSS data
have been condensed in measurements of the power spectrum at a single wave-number $k$.}
Assuming one thermalized freely-streaming sterile neutrino ($\Delta N_\nu = 1$) we find
that its mass is constrained to be $m_{\rm s} < 0.7\eV$ at $99.9\%$ C.L., a range incompatible
with the mass suggested by the LSND anomaly~\cite{LSND,review}.
%
Our results have some overlap and substantial agreement with~\cite{CMSV,MelchiorriDodelson,Seljak06,sterileothers}.




\subsection{Extra massive particles interacting among themselves}\label{sec:NnuIm}
We now assume that the extra particles discussed in section~\ref{NnuI} have a non
negligible mass $m_{\rm s}$ and are stable, such that 
when the temperature $T$ falls below $m_{\rm s}$ they form a non-relativistic relic.
(Alternatively, they could decay into freely-streaming particles
realizing a more complicated situation that is intermediate between the one studied in the previous section and in this section).
Particles with this behavior are exemplified in section~\ref{NnuI}.
In the tight coupling limit this system is described by a fluid, and the massless fluid equations in
eq.\eq{fluidevo0} generalize to
\beq\label{eq:fluidevo} \dot\delta = -(1+w) (3 \dot\Phi+k v) - 3\frac{\dot a}{a} (c_{\rm s}^2 -w) \delta,\qquad
\dot v = k\Psi -\frac{\dot a}{a}(1-3c_{\rm s}^2) v + \frac{c_{\rm s}^2}{1+w} k\delta
\eeq
where $w\equiv p/\rho$ and   $c_{\rm s}^2 \equiv \delta p/\delta \rho$ is the squared sound speed.
We fix the equation of state of the fluid by assuming that its average energy density $\rho$
and pressure $p$
is the one of
$2\cdot\Delta N_\nu$  dof decoupled from the rest of the thermal plasma:
\beq \label{eq:rhoPx} \rho = 2\cdot\Delta N_\nu\int \frac{d^3p}{(2\pi)^3} E~e^{-p/T_{\rm s}} ,\qquad
p = 2\cdot\Delta N_\nu\int \frac{d^3p}{(2\pi)^3} \frac{p^2}{3E}~e^{-p/T_{\rm s}},\qquad
E = \sqrt{p^2+m_{\rm s}^2}.\eeq
In practice this just means that the fluid interpolates between relativistic and non relativistic matter:
 $w\simeq c_{\rm s}^2 \simeq 1/3$ at $T\gg m_{\rm s}$
and $w\simeq c_{\rm s}^2\simeq 0$ at $T\ll m_{\rm s}$.
A more sophisticated treatment seems unnecessary. 
For simplicity we have adopted Boltzmann statistics in eq.\eq{rhoPx}. Again, the parameter $\Delta N_\nu$ tells the initial abundance of the extra particles in the usual `neutrino-equivalent' units. We again assume that $T_{\rm s} = T_\nu$, such that this extra component is described by two parameters: its abundance $\Delta N_\nu$ and its mass $m_{\rm s}$.

Fig.\fig{mN}b shows the result of the global fit: interacting extra particles
are constrained in a slightly stronger way than freely-streaming extra particles (at not too low C.L.).
Notice that, like in the case of massive freely-streaming sterile neutrinos, 
cosmology disfavors the mass values suggested by the LSND anoma\-ly~\cite{LSND,review}
also in this opposite limit of tightly-interacting  sterile neutrinos,
suggesting that intermediate cases might also be not viable.

A.\ de Gouvea suggested us one possible new economical interpretation of the LSND anomaly, 
in terms of decays among active-only neutrinos.
This would need some order one couplings with a light scalar, making neutrinos cosmologically interacting:
as we have seen a global fit of cosmological data disfavors this possibility.
Furthermore, a preliminary analysis indicates that the needed decay is incompatible with
SuperKamiokande atmospheric neutrino data at about $3\sigma$ C.L.

\begin{figure}[t]
$$
\includegraphics[width=0.98\textwidth]{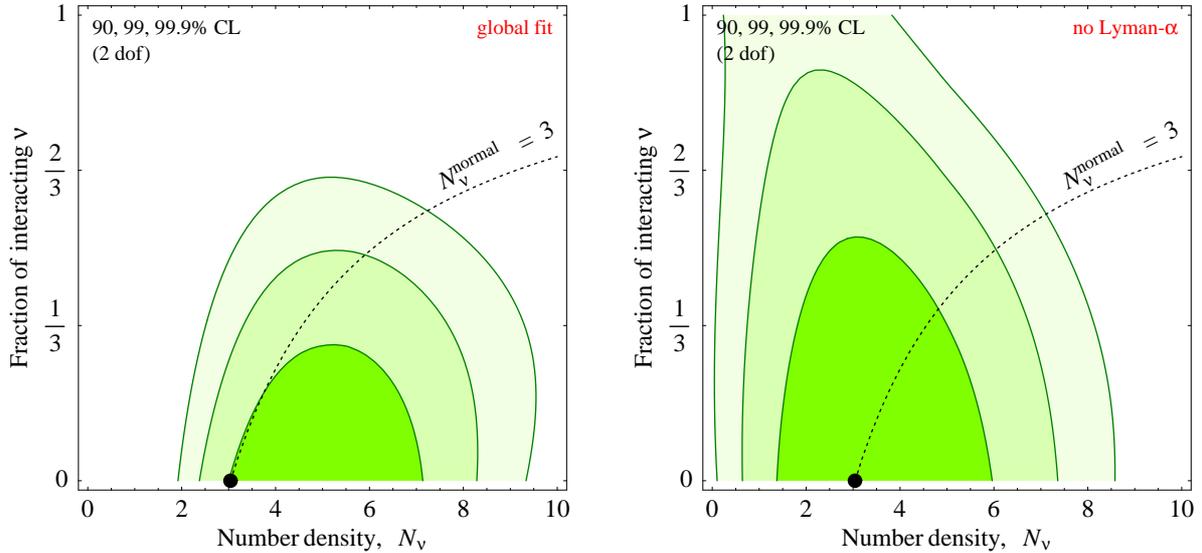}$$
\caption[X]{\label{fig:NRnu}\em Global fit 
of the total relativistic energy density
(parametrized by the usual `number of neutrinos' $N_\nu$),
and of its interacting fraction.
The dot shows the SM prediction, and
along the dotted line the number of normal freely-streaming neutrinos keeps its standard value.
}
\end{figure}

\subsection{Massless neutrinos  interacting with a massless boson}\label{00fluid}
So far we added extra light particles, free or interacting among themselves.
We now assume that ordinary neutrinos are involved in interactions with these extra particles. More specifically, we consider $N_\nu^{\rm normal}$ neutrinos that behave normally, while $N_\nu^{\rm int}$ neutrinos are involved in the interactions with extra scalars $\phi$, such that these interacting $N_\nu^{\rm int}$ neutrinos no longer free stream, but form a tightly coupled fluid together with the scalars.

Following~\cite{Hannestad,Bell}
we assume that the energy and pressure density of this fluid are given in the homogenous limit
by
\beq\rho = 2N_\nu^{\rm int} \rho_F^{\rm eq}(m_\nu, T_{\rm fl}) + N_\phi \rho_B^{\rm eq}(m_\phi,T_{\rm fl}),\qquad
p = 2N_\nu^{\rm int} p_F^{\rm eq}(m_\nu, T_{\rm fl}) + N_\phi p_B^{\rm eq}(m_\phi,T_{\rm fl})\eeq
where $\rho_F^{\rm eq}(m,T)$ and $p_F^{\rm eq}(m,T)$
are the
energy and pressure density of one fermionic degree of freedom
with mass $m$ in thermal equilibrium at temperature $T$,
and $\rho_B^{\rm eq}$ and $p_B^{\rm eq}$ are the analogous quantities for one
bosonic degree of freedom.
The temperature $T_{\rm fl}$ of the fluid is computed by assuming that 
it equals the ordinary neutrino temperature at $T\gg m_\nu,m_\phi$, and that
the fluid cools down adiabatically.
This fixes the fluid equation of state $w$ and its sound speed $c_{\rm s}$, and inhomogeneities evolve as dictated by eq.\eq{fluidevo}.
Summarizing, this system is described by the following parameters: 
$N_\nu^{\rm normal}$, $N_\nu^{\rm int}$, $N_\phi$, $m_\nu$, $m_\phi$.

\medskip

In this section we assume that $m_\nu$ and $m_\phi$ are negligibly small, such that
$w=c_{\rm s}^2 = 1/3$ (relativistic fluid).
Then, the ratio $N_\nu^{\rm int}/N_\phi$ becomes essentially irrelevant, such
that the system can be described by just two parameters: 
i) the total energy density in relativistic particles,
that we describe by the usual `number of neutrinos' 
$N_\nu = N_\nu^{\rm normal} + N_\nu^{\rm int} + 4N_\phi/7$, that remains constant;
ii) the energy fraction $R = N_\nu^{\rm int}/N_\nu$ that contributes to the fluid.
The remaining fraction $1-R =N_\nu^{\rm normal}/N_\nu$ freely streams.
In standard cosmology $R=0$ and $N_\nu = N_\nu^{\rm normal} =3.04$.
Fig.\fig{NRnu} shows how a global fit of present data determines these two parameters.
The `all interacting' case ($R=1$) is disfavored at $4\sigma$ at least 
(i.e.\ $\min \chi^2(N_\nu,R=1) -\chi^2(N_\nu=3,R=0)\circa{>}16$)
and at $3\sigma$ if  Lyman-$\alpha$ data are dropped.
As in the case of massive neutrinos, 
Lyman-$\alpha$ data make the constraint slightly stronger than the sensitivity.
Two previous analyses claimed different results:
our constraints are somewhat stronger than in~\cite{Bell} (possibly because we use the most recent data set)
and weaker than in~\cite{Hannestad}.

\begin{figure}[t]
$$\includegraphics[width=0.99\textwidth]{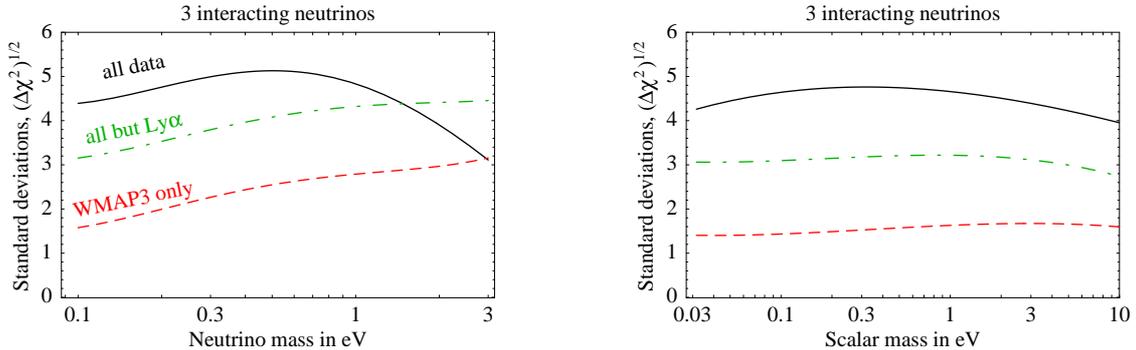}$$
\caption{\label{fig:chiplot}\em Number of standard deviations,
defined as $(\chi^2 - \chi^2_{0})^{1/2}$
(where $\chi^2_0$ is the best  $\Lambda\rm CDM$ fit with massless neutrinos)
at which cosmological data disfavor a fluid of 3 neutrinos interacting with a scalar
assuming massive neutrinos and massless scalar (fig.\fig{chiplot}a) or massless neutrinos and massive scalar (fig.\fig{chiplot}b).
The different lines correspond to different data-sets: global fit (continuous black line),
Lyman-$\alpha$ data dropped (dot-dashed green line), WMAP3 only (dashed red line).}
\end{figure}

\subsection{Massive neutrinos  interacting with a massless boson}\label{m0fluid}
We now explore how the situation changes if neutrinos have a non vanishing mass $m_\nu$.
We focus on the most interesting limiting case:  $R=0$ i.e.\
we now assume that all neutrinos are involved in the interaction.
This is interesting because it means that the cosmological bound on neutrino masses 
no longer applies, because
when $T\circa{<} m_\nu$ all neutrinos annihilate or decay into massless $\phi$ particles.
Scenarios of this kind have been proposed for a number of reasons~\cite{Nuless,LateMasses,MaVaNs-like}.
We again assume that neutrinos initially have the standard abundance, and that bosons initially 
have the minimal
abundance, $N_\phi =1$ (one real scalar).
After that all neutrinos annihilate into $\phi$, they acquire a relativistic energy density
corresponding to an equivalent number of neutrinos $N_\nu(T\circa{<} m_\nu) = 4/7 (25/4)^{4/3} \sim 6.6$.

Fig.\fig{chiplot}a shows how much this
non-standard cosmology is disfavored as a function of $m_\nu$
(standard cosmology is not recovered for any value of $m_\nu$).
For $m_\nu \ll \eV$ the result is similar to the case $m_\nu=0$, already discussed in section~\ref{00fluid}: this scenario is disfavored at about $4\sigma$
by the global fit.
As already noticed in~\cite{Bell}, the scenario becomes less disfavored for  $m_\nu\circa{>}\eV$
 (beta decay data demand $m_\nu\circa{<}2\eV$~\cite{review}).
 We find that WMAP3 data (dashed lines in fig.s\fig{chiplot})
  are more constraining than the WMAP1 data analyzed in~\cite{Bell}.

We do not consider intermediate scenarios where only one or two massive neutrinos interact with the scalar: both the constraint on neutrino masses and on their  free-streaming  applies,
but in a milder form~\cite{Bell}.

\subsection{Massless neutrinos  interacting with a massive boson}\label{0mfluid}
We conclude studying the opposite limit: neutrinos have a negligibly small mass,
while the scalar has a mass $m_\phi$.
The considerations of the previous section still apply, with the r\^ole of neutrinos
and interacting particles interchanged: since neutrinos have more degrees of freedom than one scalar, 
the radiation density increases in a mild way when $T$ drops below $m_\phi$.
Fig.\fig{chiplot}b shows how much this
non-standard cosmology is disfavored as a function of $m_\phi$: we find
almost no dependence on $m_\phi$: this scenario is disfavored at about $4\sigma$.
For large enough $m_\phi$, depending on the model,
interactions mediated by $\phi$ must become weak enough that
neutrinos recover their standard freely-streaming behavior.

\section{Conclusions}

We compared  a non exhaustive but representative casistics of
how cosmology is affected by extra light particles (with sub-keV masses),
or by standard and non-standard properties of neutrinos, using CMB, LSS, Lyman-$\alpha$, BAO, SN data. 
\begin{itemize}
\item 
First, we considered ordinary massive neutrinos.
We obtain the cosmological bound on neutrino masses, $\sum m_\nu \circa{<}0.40\eV$
at $99.9\%$ C.L.\
and fig.\fig{Nm}a shows that the relatively less safe observations play a crucial r\^ole.
\item 
The density of initially relativistic particles can be parameterized 
in terms of the usual number $N_\nu$ of equivalent neutrinos.
Assuming that all the $N_\nu$ relativistic particles freely stream, we find that their density is constrained to be
$N_\nu=5\pm 1$. 
The $2\sigma$ preference for $N_\nu>3$ is mainly due to the $2\sigma$ anomaly
in the Lyman-$\alpha$ measurement of the matter power spectrum.

\item Assuming ordinary neutrinos plus an extra component of interacting particles,
we find $\Delta N_\nu = 0\pm1.3$.
Fig.\fig{NRnu} shows how data constrain the intermediate case where both kinds of relativistic
particles are present.
It is interesting that the uncertainty on $\Delta N_\nu$ is decreasing below 1.

\item 
The extra light particles might have a mass $m$ and an abundance $\Delta N_\nu$.
Fig.\fig{mN} shows how data constrain these parameters in the two limiting cases that these extra particles freely stream 
(fig.\fig{mN}a) or interact among themselves (fig.\fig{mN}b).
\item 
Finally, we considered one extra scalar of mass $m_\phi$ that interacts with neutrinos of mass $m_\nu$.
We find that this scenario is strongly disfavored by the global fit, at about $4\sigma$.
\end{itemize}
All these results are based on assumptions and subject to caveats, that we discussed in the text.
Technically, our analysis somewhat differs from typical analyses because
we used a code developed by us and dealt with statistics using Gaussian analytical techniques,
that become adequate nowadays that  observations are rich and precise enough.
Eq.s \eq{means} and\eq{corr} allow to check how well we reproduce the 
standard results for standard cosmology.

\small

\paragraph{Acknowledgments}
We thank Ben Allanach, Paolo Cr{\em e}minelli, Concha Gonzalez-Garcia, Danilo Marchesini, Riccardo Rattazzi, Adam Riess and Licia Verde for discussions and clarifications and 
John Kovac, Hiranya Peiris and Jonathan Sievers for communications.
The authors thank the Galileo Galilei Institute for Theoretical Physics in Firenze for the hospitality and INFN for partial support during the completion of this work.
The work of M.C. is supported in part by the USA DOE-HEP Grant DE-FG02-92ER-40704.  
M.C. thanks the Service de Physique Th\'eorique of CEA-Saclay for the hospitality during the completion of this work. 

\begin{multicols}{2}

\end{multicols}

\end{document}